\documentclass[preprint,12pt]{elsarticle}
\usepackage{amssymb,amsmath,epsf}
\usepackage{array}

\journal{Nuclear Physics A}

\def\p{{\mbox{\boldmath$p$}}}
\def\tlab{\mbox{$T_{\rm Lab}$}}

\def\vz{{\mbox{\boldmath$0$}}}

\def\pzp{p_0^{\prime}}

\newcommand{\bp}{{\bf p}}
\newcommand{\bpp}{{\bf p}^{\prime}}
\newcommand{\bk}{{\bf k}}

\def\Pto{^3P_{1}^{o}}

\def\Sp{^1S_{0}^{+}}

\def\Sp{^1S_0}
\def\Ps{^3P_0}
\def\Pto{^1P_1}
\def\Ptt{^3P_1}

\begin{document}

\begin{frontmatter}

\title{Relativistic multirank interaction kernels of the neutron-proton system}

\author[jinr]{S.\,G. Bondarenko}
\ead{bondaren@jinr.ru}
\author[jinr]{V.\,V. Burov}
\ead{burov@theor.jinr.ru}
\author[hwa]{W.-Y. Pauchy Hwang}
\ead{wyhwang@phys.ntu.edu.tw}
\author[jinr]{E.\,P. Rogochaya\corref{cor1}}
\ead{rogoch@theor.jinr.ru}
\address[jinr]{Bogoliubov Laboratory of Theoretical Physics, Joint Institute for Nuclear Research, Dubna, Russia}
\address[hwa]{National Taipei University, Taipei 106, Taiwan}

\cortext[cor1]{Corresponding author. JINR, Joliot-Curie 6, 141980
Dubna, Moscow region, Russia. Tel.: +74962163503; fax: +74962165146}

\begin{abstract}
The multirank separable kernels of the neutron-proton interaction
for uncoupled $S$ and $P$ partial waves (with the total angular
momentum $J$=0,1) are proposed. Two different methods of a
relativistic generalization of initially nonrelativistic form
factors parametrizing the kernel are considered. Using the
constructed kernels the experimental data for phase shifts in the
elastic neutron-proton scattering for the laboratory energy up to
3 GeV and low-energy parameters are described. The comparison of
our results with other model calculations are presented.
\end{abstract}


\end{frontmatter}

\section{Introduction}
The problem of an adequate description of nuclear interactions
arose many years ago. An important role in this task is played by
the construction of the nucleon-nucleon interaction. The simplest
way to investigate such interaction is to describe properties of
the elastic neutron-proton ($np$) scattering and their bound state
- deuteron. The latter can be considered through some reactions,
such as the photodisintegration, the electrodisintegration etc.

There are a lot of works devoted to the description of the
deuteron. The first models were based on the nonrelativistic
Shr$\ddot{\rm o}$dinger equation (see, for example,
\cite{Brown:1975us,Mathiot:1984bs}). In numerous papers mesonic
exchange currents, relativistic corrections were investigated and
then were added to nonrelativistic solutions
\cite{Gari:1976kj}-\cite{Carbonell:1998rj}.
These approaches could describe properties of the deuteron such as
binding energy, magnetic and quadruple momenta, electromagnetic
form factors, tensor polarization and so on. However, with
increasing of the precision of experimental data and obtaining new
data at larger energies it became evident that it was necessary to
take into account relativistic effects more carefully. And the
consistency of the consideration of the deuteron breakup reactions
demands also the final state interaction (FSI) between the
outgoing nucleons to be taken into account.

One of the most consistent approaches is based on the solution of
the Bethe-Salpeter (BS) equation \cite{Salpeter:1951sz}. In this
case, we have to deal with a nontrivial integral equation. In
addition to introduce the FSI of the final $np$ pair the BS
equation for the continuous state should be solved. There is no
method to get its exact solution. So various approximations were
worked out. At present, the most known and developed approaches
are the so-called quasipotential
\cite{Logunov:1963yc}-\cite{Garcon:2001sz}.

They consist in the simplification of the equations under
consideration by some assumption. In most cases it means the
deliverance from the relative energy, according to some physical
reasons. For example, one of the particles is supposed
on-mass-shell \cite{Adam:2002cn}, or the time coordinates of
particles are made equal \cite{Pascalutsa:2000bs}, etc. There is
another approach called the light front (LF) dynamics which was
successfully developed and applied to explain the electromagnetic
properties of the deuteron. In this approach, the state vector
describing the system is expanded by Fock components defined on a
hypersphere in the four-dimensional space-time. The LF dynamics
approach is intuitively appealing, since it is formally close to
the nonrelativistic description in terms of the Hamiltonian and
state vectors can be directly interpreted as wave functions (see,
for example, review \cite{Carbonell:1998rj}). The equivalence
between LF dynamics and BS approaches was a subject of discussions
presented in \cite{Carbonell:1998rj,Bondarenko:2002zz} and
references therein.

An alternative approach based on the exact solution of the BS
equation is to use the separable ansatz for the interaction kernel
in the BS equation \cite{Bondarenko:2002zz}. In this case we can
transform an integral equation to a system of linear equations.
Parameters of the kernel are fitted by the description of phase
shifts for respective partial states and low-energy parameters.

First separable parametrizations were worked out within
nonrelativistic models. The form factors in the interaction kernel
had no poles on a real axis in the relative energy complex plane
\cite{Mathelitsch:1981mr,Haidenbauer:1984dz}. However, after the
construction of a relativistic generalization such poles appeared
\cite{Bondarenko:2002zz}. In some cases they do not prevent to
perform the calculations. As an example, such parametrizations are
successfully used in the consideration of the deuteron
photodisintegration \cite{Bondarenko:2004pn}, elastic
electron-deuteron scattering, deuteron electrodisintegration on
the threshold, and the deep inelastic scattering
\cite{Bondarenko:2002zz}.
However, at high energies, one would have to deal with several
thresholds corresponding to the production of one, two and more
mesons of different types. This is clearly not feasible. A more
practical approach is to employ phenomenological covariant
separable kernel, which do not exhibit the meson-production
thresholds, and can even be constructed in a singularity-free
fashion, with the form factors chosen in the present paper and our
Wick-rotation prescription. Thus, an accurate description of
on-shell nucleon-nucleon data is possible, up to quite high
energies. One then hopes that the used separable interactions also
have a reasonable off-shell behaviour, so that realistic
applications to other reactions can be done.
The parametrization like that was proposed in
\cite{Schwarz:1980bc}. In our previous works
\cite{Bondarenko:2008fp,Bondarenko:2008ha}, we constructed the
one-rank interaction kernel of the same type. Due to the
simplicity it works till the laboratory energy $\tlab\sim$1\,GeV
except the simplest partial wave $^1P_1^+$ where it gives
satisfactory results for all available experimental data. In the
present work, we develop this approach increasing the rank of the
kernels and trying to describe the data for the phase shifts of
uncoupled partial states in the energy range up to 3\,GeV
taken from the SAID program (http://gwdac.phys.gwu.edu). In future
we plan to use these kernels in relativistic calculations of the
deuteron electrodisintegration far from the threshold.

It should be emphasized that in this paper the new kernel is
fitted to describe the neutron-proton elastic scattering data
only. The description of the proton-proton scattering is a
separate problem which requires specific methods of calculation.
Our aim was to construct the separable kernels suitable for
consideration of the scattered $\Pto,\Ps,\Ptt,\Sp$ states of the
neutron-proton system.

The paper is organized as follows. In Section \ref{sect2}, the
general Bethe-Salpeter formalism is considered. The used separable
kernel is described in Section \ref{sect3}. Section \ref{sect4} is
devoted to the methods of a relativistic generalization of
nonrelativistic Yamaguchi- and Tabakin-type form factors. In
Section \ref{sect5}, the pole structure of the obtained
relativistic expressions is analyzed. The parametrizations for
definite partial channels are presented in Section \ref{sect6}. In
Section \ref{sect7}, the scheme of performing numerical
calculations is offered. The obtained results and the comparison
with other model calculations are discussed in Section
\ref{sect8}. In conclusion, in Section \ref{sect9}, the fields of
application of the constructed kernels is briefly outlined.
%
\section{Bethe-Salpeter formalism}\label{sect2}
Within the relativistic field theory, the elastic NN scattering
can be described by the scattering $T$ matrix which satisfies the
inhomogeneous BS equation. In momentum space, the BS equation for
the $T$ matrix can be (in terms of the relative four-momenta
$p^\prime$ and $p$ and the total four-momentum $P$) represented as:
\begin{eqnarray}
T(p^{\prime}, p; P) = V(p^{\prime}, p; P) + \frac{i}{4\pi^3}\int
d^4k\, V(p^{\prime}, k; P)\, S_2(k; P)\, T(k, p; P), \label{t00}
\end{eqnarray}
where $V(p^{\prime}, p; P)$ is the interaction kernel and
$S_2(k; P)$ is the free two-particle Green function
$$S_2^{-1}(k; P)=\bigl(\tfrac12\:P\cdot\gamma+{k\cdot\gamma}-m\bigr)^{(1)}
\bigl(\tfrac12\:P\cdot\gamma-{k\cdot\gamma}-m\bigr)^{(2)},$$
$\gamma$ are the Dirac gamma-matrices.
The square of the total momenta $s=(p_1+p_2)^2$ and the
relative momentum $p=(p_1-p_2)/2$ [$p'=(p_1'-p_2')/2$] are
defined via the nucleon momenta
$p_1,~p_2$ [$p_1',~p_2'$] of initial [final] nucleons.

To perform the partial-wave decomposition of the BS equation (\ref{t00}),
we introduce relativistic two-nucleon basis states
$|aM\rangle\equiv|\pi,\,{}^{2S+1}L_J^{\rho} M \rangle$, where $S$
denotes the total spin, $L$ is the orbital angular momentum, and $J$
is the total angular momentum with the projection $M$;
relativistic quantum numbers $\rho$ and $\pi$ refer to the
relative-energy and spatial parity
with respect to the change of sign of the
relative energy and spatial vector, respectively.
Then the partial-wave decomposition
of the $T$ matrix in the center-of-mass frame (c.m.) has the following form:
\begin{eqnarray}
&&T_{\alpha\beta,\gamma\delta}(p^{\prime},p; {P_{(0)}}) \nonumber\\
&&= \sum_{JMab}
({\cal Y}_{aM}(-{\bpp})U_C)_{\alpha\beta}\otimes (U_C {\cal
Y}^{\dag}_{bM}({\bp}))_{\delta\gamma}\
T_{ab}(\pzp,|\bpp|;p_0,|\bp|;s), \label{t01}
\end{eqnarray}
where $U_C =i\gamma^2\gamma^0$ is the charge conjugation matrix;
the total momentum of the colliding nucleons in c.m. is denoted by $P_{(0)}$.
Greek letters $(\alpha,\beta)$ and $(\gamma,\delta)$ in
Eq.\eqref{t01} refer to spinor indices and label particles in the
initial and final states, respectively. It is convenient to
represent the two-particle states in terms of matrices. To this
end, the Dirac spinors of the second nucleon are transposed. At
this stage $T$ is $16\times16$ matrix in spinor space which,
sandwiched between Dirac spinors and traced, yields the
corresponding transition matrix elements between $SLJ$-states.

The spin-angular momentum functions ${\cal Y}_{aM}({\bp})$ are
expressed in terms of the positive- and negative-energy Dirac
spinors $u^{\rho=\pm,\, 1/2}_{m}$, the spherical harmonics
$Y_{L{m_L}}$ and Clebsch-Gordan coefficients $C_{j_1 m_1 j_2
m_2}^{j\:m}$:
\begin{eqnarray}
{\cal Y}_{JM:LS {\rho}}(\bp) U_C&&
\nonumber\\
= && i^{L} \sum_{m_Lm_Sm_1m_2\rho_1\rho_2}C_{\frac12 \rho_1 \frac12
\rho_2}^{S_{\rho} {\rho}} C_{L m_L S m_S}^{JM} C_{\frac12 m_1
\frac12 m_2}^{Sm_S} \times\nonumber\\
&&~~~~~~~~~~~~~~~~~Y_{L{m_L}}(\bp) {u^{\rho_1}_{m_1}}^{(1)}(\bp)
{{u^{\rho_2}_{m_2}}^{(2)}}^{T}(-\bp). \label{s01}
\end{eqnarray}
The superscripts in Eq.\eqref{s01} refer to particles (1) and (2).
To derive the matrix elements between $a$-states, the
ortonormalization condition for the functions ${\cal
Y}_{aM}({\bpp})$ should be used:
\begin{eqnarray}
\int\!d\varphi_{\bp}\:d(\cos\theta_{\bp})\,{\rm Tr}\left\{ {\cal
Y}^{\dag}_{aM}({\bp}){\cal
Y}_{a^{\prime}M^{\prime}}({\bp})\right\}
\hskip 60mm\nonumber\\
\equiv \int\!d\varphi_{\bp}\:d(\cos\theta_{\bp}) ({\cal
Y}^{\dag}_{aM}({\bp}))_{\beta\alpha} ({\cal
Y}_{a^{\prime}M^{\prime}} ({\bp}))_{\alpha\beta} =
\delta_{aa^{\prime}}\delta_{MM^{\prime}}, \label{t02}
\end{eqnarray}
where partial states $a$ and $a^{\prime}$ belong to the same partial channel.

The partial-wave decomposition for the interaction kernel $V$ of
the BS equation~\eqref{t00}  can be written analogously to
Eq.\eqref{t01}:
\begin{eqnarray}
&&V_{\alpha\beta,\gamma\delta}(p^{\prime},p; P_{(0)}) \nonumber\\
&&= \sum_{abM}
({\cal Y}_{aM}(-{\bpp})U_C)_{\alpha\beta}\otimes (U_C{\cal
Y}^{\dag}_{bM}({\bp}))_{\delta\gamma}\
V_{ab}(\pzp,|\bpp|;p_0,|\bp|;s). \label{t03}
\end{eqnarray}
Applying the condition~\eqref{t02}, we can obtain a system of
linear integral equations for the off-shell partial-wave
amplitudes:
\begin{eqnarray}
&&T_{ab}(\pzp, |\bpp|; p_0, |\bp|; s) =
V_{ab}(\pzp, |\bpp|; p_0, |\bp|; s)
\hskip 50mm
\label{t04}\\
&&\hskip 38mm+ \frac{i}{4\pi^3}\sum_{cd}\int\limits_{-\infty}^{+\infty}\!
dk_0\int\limits_0^\infty\! \bk^2 d|\bk|\, V_{ac}(\pzp, |\bpp|; k_0,
|\bk|; s)\nonumber\\
&&\hskip 38mm\times\, S_{cd}(k_0,|\bk|;s)\, T_{db}(k_0,|\bk|;p_0,|\bp|;s), \nonumber
\end{eqnarray}
where the two-particle propagator $S_{ab}$ depends only on
$\rho$-spin indices.

We use the normalization condition for the $T$ matrix in the
on-mass-shell form for the singlet case:
\begin{eqnarray}
T_{ll}(s)\equiv T_{ll}(0,\bar p;0, \bar p;s)=-\frac{16\pi}{\sqrt
s\sqrt{s-4m^2}}\exp\{i\delta_l\}\sin\delta_l, \label{T_norm_s}
\end{eqnarray}
where $\bar p\equiv |{\bar \bp|} =
\sqrt{s/4-m^2}=\sqrt{m\tlab/2}$ is the on-mass-shell
momentum, $m$ is a nucleon mass.
In Eq.(\ref{T_norm_s}) $l$ denotes
$^SL_J$ states for simplicity. Low-energy parameters, the scattering
length $a_0$, and the effective range $r_0$ are derived from the
expansion of the $T$-matrix into a series of $\bar p$-terms,
according to \cite{Bethe:1949yr}:
\begin{eqnarray}
\bar p\cot\delta_l(s)=-\frac{1}{a_0^l}+\frac{r_0^l}{2}\bar p^2
+{\cal O}(\bar p^3).\label{low}
\end{eqnarray}

To solve the equations for the $T$ matrix and BS amplitude, we should
use some assumption for the interaction kernel.

\section{A separable kernel}\label{sect3}

We assume that the interaction kernel $V$ conserves parity,
total angular momentum  $J$ and its projection, and isotopic spin.
Due to the tensor nuclear force, the orbital angular momentum
$L$ is not conserved. Moreover, the negative-energy two-nucleon
states are switched off, which leads to the total spin $S$
conservation. The partial-wave-decomposed BS equation is therefore
reduced to the following form:
\begin{eqnarray}
T_{l'l}(\pzp, |\bpp|; p_0, |\bp|; s) =
V_{l'l}(\pzp, |\bpp|; p_0, |\bp|; s)
\hskip 50mm
\label{t05}\\
+ \frac{i}{4\pi^3}\sum_{l''}\int\limits_{-\infty}^{+\infty}\!
dk_0\int\limits_0^\infty\! \bk^2 d|\bk|\, \frac{V_{l'l''}(\pzp,
|\bpp|; k_0,|\bk|; s)\, T_{l''l}(k_0,|\bk|;p_0,|\bp|;s)}
{(\sqrt{s}/2-E_{\bk}+i\epsilon)^2-k_0^2}, \nonumber \label{BS}
\end{eqnarray}
where $l=l^{\prime}=J$ for spin-singlet and uncoupled spin-triplet
states.

Supposing the separable (rank $N$) ansatz for the kernel of the NN
interaction:
\begin{eqnarray}
V_{l'l}(\pzp, |\bpp|; p_0, |\bp|; s)=\sum_{i,j=1}^N\lambda_{ij}(s)
g_i^{[l']}(\pzp, |\bpp|)g_j^{[l]}(p_0, |\bp|), \label{V_separ}
\end{eqnarray}
where the form factors $g_j^{[l]}$ represent the model functions, we can
obtain the solution of equation (\ref{BS}) in a similar
separable form for the $T$ matrix:
\begin{eqnarray}
T_{l'l}(\pzp, |\bpp|; p_0, |\bp|; s)=
\sum_{i,j=1}^N\tau_{ij}(s)g_i^{[l']}(\pzp, |\bpp|)
g_j^{[l]}(p_0, |\bp|),
\end{eqnarray}
where
\begin{eqnarray}
\tau_{ij}(s)=1/(\lambda_{ij}^{-1}(s)+h_{ij}(s)),
\end{eqnarray}
\begin{eqnarray}
h_{ij}(s)=-\frac{i}{4\pi^3}\sum_{l}\int dk_0\int
\bk^2d|\bk| \frac{g_i^{[l]}(k_0,|\bk|)g_j^{[l]}(k_0,|\bk|)}{(\sqrt
s/2-E_{\bk}+i\epsilon)^2-k_0^2}, \label{H_separ}
\end{eqnarray}
$\lambda_{ij}(s)$ is a matrix of model parameters.

The form factors $g_i^{[l]}$ used in the separable representation of the
interaction kernel (\ref{V_separ}) are obtanied by a relativistic
generalization of the initially nonrelativistic Yamaguchi-type
functions depending on the three-dimensional squared momentum $|\p|$.
There are two methods to derive covariant relativistic
generalizations of nonrelativistic form factors. They are
considered in the next section.

Calculating the $T$ matrix we can connect the parameters of the internal kernel
with observables.

\section{Methods of a covariant relativistic generalization}\label{sect4}
In this section, two methods of a covariant relativistic
generalization of the Yamaguchi- and Tabakin-type functions
are presented.
\begin{enumerate}
\item One of the common methods is to
replace three-momentum squared by four-momentum squared:
\begin{eqnarray}
\bp^2 \to -p^2 = -p_0^2 + \bp^2. \label{p2p}
\end{eqnarray}
This formal procedure converts three-dimensional functions to
covariant four-dimensional ones. \item The other method is based on the
introduction of the formal four-vector $Q$ via the relative $p$
and total $P$ four-momenta of the two-body system by the following
relation:
\begin{eqnarray}
Q = p - \frac{P\cdot p}{s} P, \label{Q2p}
\end{eqnarray}
with the total momentum squared $s=P^2$.
\end{enumerate}
Note that in the two-particle center-of-mass system where
$P=(\sqrt{s},\vz)$ the four-vector $Q$ is defined by the components
$Q=(0,\bp)$ and, thus,
\begin{eqnarray}
\bp^2 = -Q^2 \label{p2Q}
\end{eqnarray}
can be formally converted to the Lorentz invariant.

Let us consider the methods described above as applied to the
nonrelativistic Yamaguchi-type function
\begin{eqnarray}
g(|\bp|) = \frac{1}{\bp^2+\beta^2}. \label{yams}
\end{eqnarray}

In the first case, using the substitution (\ref{p2p}) we obtain the
covariant function in the form:
\begin{eqnarray}
g_p(p,P) = \frac{1}{-p^2+\beta^2} \stackrel{\rm
c.m.}{\longrightarrow}
\frac{1}{-p_0^2+\bp^2+\beta^2+i\epsilon}. \label{myams}
\end{eqnarray}

In the second case, we use relation (\ref{p2Q}) and obtain the
function:
\begin{eqnarray}
g_Q(p,P) = \frac{1}{-Q^2+\beta^2} \stackrel{\rm
c.m.}{\longrightarrow} \frac{1}{\bp^2+\beta^2}. \label{myamQs}
\end{eqnarray}

The presented functions have rather different properties in the
relative energy $p_0$ complex plane in c.m. The function $g_p$ has
two poles on the real axis for $p_0$ at $\pm \sqrt{\bp^2+\beta^2}
\mp i\epsilon$ while the function $g_Q$ has no poles on it.

In practical calculations of the reactions with the high momentum
transfer the $p_0$ integration can lead to singular expressions in
functions $g_{p,Q}$ on $|\bp|$ or $\cos{\theta_{\bp}}$. This
problem can be easily solved by calculating the $|\bp|$ or
$\cos{\theta_{\bp}}$ principal value integral. However, another
form of functions $g_{p,Q}$ with odd powers in the denominator leads
to nonintegrable singularities. Therefore, we introduce functions
$g_{p,Q}$ of type without poles on the real axis in the relative
energy $p_0$ complex plane. As an example of a function like that we
introduce the covariant form factors in the following form (see
also section 3 of \cite{Bondarenko:2008fp}):
\begin{eqnarray}
g_p(p,P) = \frac{p_c-p^2}{(p^2-\beta^2)^2+\alpha^4} \stackrel{\rm
c.m.}{\longrightarrow}
\frac{p_c-p_0^2+\bp^2}{(p_0^2-\bp^2-\beta^2)^2+\alpha^4},
\label{myamsa}
\end{eqnarray}
and in the second case we use relation (\ref{p2Q}) and obtain
the function
\begin{eqnarray}
g_Q(p,P) = \frac{p_c-Q^2}{(Q^2-\beta^2)^2+\alpha^4} \stackrel{\rm
c.m.}{\longrightarrow}
\frac{p_c+\bp^2}{(\bp^2+\beta^2)^2+\alpha^4}. \label{myamQsa}
\end{eqnarray}
We note that the function $g_Q$ still has no poles on the $p_0$
real axis while $g_p$ has poles at $p_0$: $\pm
\sqrt{\bp^2+\beta^2 + i\alpha^2},\quad \pm \sqrt{\bp^2+\beta^2 -
i\alpha^2}$.

The two methods of a covariant relativistic generalization
described above can be investigated by solving the Bethe-Salpeter
equation for specific partial states.

\section{Pole structure of the BS solution}\label{sect5}

The solution of the Bethe-Salpeter equation with the separable
kernel of interaction contains the function $h_{ij}$
(\ref{H_separ}). To simplify the investigation of the pole
structure, let us consider $h_{ij}$ for the one-rank kernel
($i=j=1$) for single $l$ state. Then the value to be calculated is
$h(s)$:
\begin{eqnarray}
h(s)=-\frac{i}{4\pi^3}\int dp_0\int\bp^2d|\bp|
\frac{{g(p_0,|\bp|)^2}}{{(\sqrt{s}/2-E_{\bp}+i\epsilon)^2-p_0^2},}. \label{hs}
\end{eqnarray}

To obtain the function $h(s)$, the two-dimensional integral on
$p_0$ and $|\bp|$ should be calculated. To perform the integration
over $p_0$, the Cauchy theorem is used. As it can be seen from
Eq.(\ref{hs}), there are two types of singularities on the real
axis in the $p_0$ complex plane: one is poles of the
function $S(p;s)$:
\begin{eqnarray}
p^{(1,2)}_0 = \pm \sqrt{s}/2 \mp E_{\bp} \pm i\epsilon,
\label{p0a1}
\end{eqnarray}
and the other is poles of the function $g(p)$.

\begin{figure}
\begin{center}
\includegraphics[width=95mm]{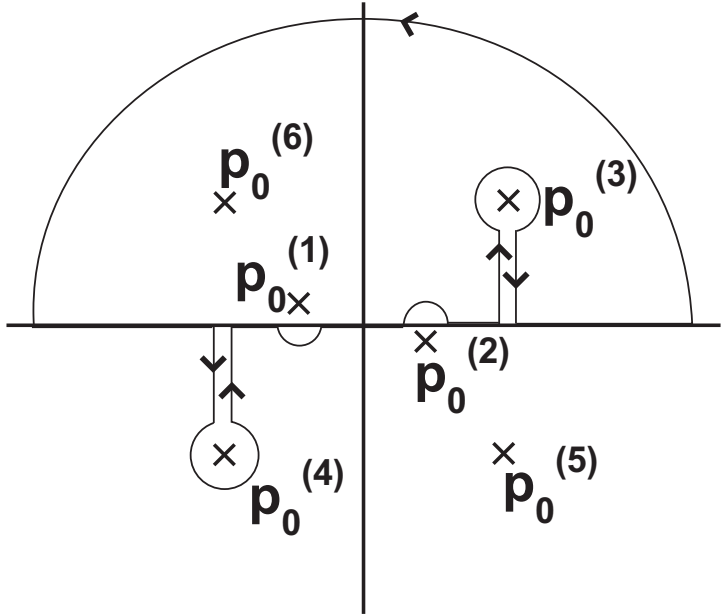}
\caption{{Contour for integration over $p_0$.}}
\label{contour}
\end{center}
\end{figure}

The function $g_p$ has four poles:
\begin{eqnarray}
p^{(3,4)}_0 = \pm \sqrt{\bp^2+\beta^2 + i\alpha^2},
\nonumber\\
p^{(5,6)}_0 = \pm \sqrt{\bp^2+\beta^2 - i\alpha^2}, \label{p0a2}
\end{eqnarray}
and to perform the $p_0$ integration, residues in three poles of
Eqs.(\ref{p0a1}) and (\ref{p0a2}) should be calculated. These
calculations are performed analytically. This procedure is worthy
of a special discussion. All poles and the contour of integration
are pictured in Fig.\ref{contour}. The idea how to choose the
contour appeared owing to \cite{Cutkosky:1969fq,Lee:1969fy}. It
consists in that the contour must envelope the poles from form
factors which will be inside the standard contour after the
$\alpha\rightarrow 0$ limit. "Standard"\ means the one used in the
quantum field theory calculations with a propagator which has
poles only on the real axis in the $p_0$ complex plane; one of
them is rounded from below and the other, from above. So the path
of integration is defined by an appropriate contour for the
propagator. The calculation over the presented path leads to the
pure real contribution from the form factor poles and, therefore,
to the unitary $T$ matrix. We also obtain a correct transition to
ordinary form factors of type $g\sim 1/(p_0^2-\bp^2-\beta^2)^2$ in
the $\alpha\rightarrow 0$ limit.

The function $g_Q$ has no poles on the $p_0$ real axis and,
therefore, the only poles of Eq.(\ref{p0a1}) should be taken into
account. The result for $h(s)$ can be written as:
\begin{eqnarray}
h(s)=\frac{1}{2\pi^2}\int \bp^2d|\bp|
\frac{g_Q(0,|\bp|)^2}{\sqrt s-2E_{\bp}+i\epsilon}. \label{hsp}
\end{eqnarray}

This equation formally coincides with that could be obtained
within the Blankenbeckeler-Sugar-Logunov-Tavkhelidze (BSLT)
approximation~\cite{Logunov:1963yc,Blankenbecler:1965gx}
which consists in replacing the Green
function in Eq.(\ref{hs}) by the expression
\begin{eqnarray}
S_{BSLT}(p;s) = -2\pi i (\sqrt s-2E_{\bp}+i\epsilon)^{-1}
\delta(p_0). \label{sgrbslt}
\end{eqnarray}

Although the solutions of the equation with functions $g_Q$ and
within the BSLT approximation coincide in c.m., the difference
becomes evident when the reaction with the two-particle system is
considered. In that case, the arguments of the function $g_Q$ are
calculated with the help of the Lorentz transformations in the
system different from c.m. The functions become similar to $g_p$
but with a more complicated dependence of the $p$ argument. In our
calculations we prefer to use the first method of relativization.
The reason is in that reaction we are planning to consider in
future (namely, electrodisintegration) these transformations could
lead to the appearance of additional poles in the calculated
expressions. However, in this work the comparison of results for phase
shifts and low-energy parameters is presented for both cases.

In following two sections some separable presentations of the
interaction kernel for the partial waves with $J=0,1$ are
considered. Form factors are constructed by the relativization
procedure, according to the first method. The functions with $Q$ can
be obtained from them by the change $p^2\rightarrow Q^2$.

\section{Separable presentations of the kernel}\label{sect6}
We consider the partial states using the separable kernels
with modified Yamaguchi-type functions of $N$ rank (MY$N$, MYQ$N$).
\subsection{Two-rank kernel for $P$ states: $\Ps^+$, $\Pto^+$, $\Ptt^+$}
For the description of the $P$ partial waves the two-rank
separable kernel of interaction with the modified
Yamaguchi-type functions is used. Its form factors
are written as:
\begin{eqnarray}
&&g^{[P]}_1(p)=\frac{\sqrt{-p_0^2+\bp^2}}
{(p_0^2-\bp^2-\beta_{1}^2)^2+\alpha_{1}^4},
\\
&&g^{[P]}_2(p)=\frac{\sqrt{(-p_0^2+\bp^2)^3}(p_{c2}-p_0^2+\bp^2)}
{((p_0^2-\bp^2-\beta_{2}^2)^2+\alpha_{2}^4)^2}.
\nonumber
\end{eqnarray}

\subsection{Three-rank kernel for $\Sp^+$ state}
The investigation we performed demonstrates a bad description of
phase shifts for $^1S_0^+$ partial state by the two-rank kernel.
It is not much better than by the one-rank kernel \cite{Bondarenko:2008fp}. So
for this special case the three-rank interaction kernel was
elaborated. Its form factors have the following form:
\begin{eqnarray}
&&g^{[S]}_1(p)=\frac{(p_{c1}-p_0^2+\bp^2)}
{(p_0^2-\bp^2-\beta_{1}^2)^2+\alpha_{1}^4},
\\
&&g^{[S]}_2(p)=\frac{(p_0^2-\bp^2)(p_{c2}-p_0^2+\bp^2)^2}
{((p_0^2-\bp^2-\beta_{2}^2)^2+\alpha_{2}^4)^2},
\nonumber\\
&&g^{[S]}_3(p)=\frac{(p_0^2-\bp^2)}
{(p_0^2-\bp^2-\beta_{3}^2)^2+\alpha_{3}^4}.
\nonumber
\end{eqnarray}
The functions are numerated by angular momenta $L=0([S]),1([P])$.
The numerator with $p_c$ in $g_2^{[P]}$ and $g^{[S]}_{1,2}$ is
introduced to compensate an additional dimension in the
denominator to provide the total dimension as GeV$^{-2}$
\cite{Bondarenko:2008fp}.
\section{Calculations and results}\label{sect7}
Using the $np$ scattering data we analyze the parameters
of the separable kernels
distinguishing three different cases:
\begin{enumerate}
\item There are no sign change in phase shifts or bound state
($^1P_1^+,~^3P_1^+$ partial states). In this case
\begin{eqnarray}
\lambda_{ij}(s)=\bar\lambda_{ij}=const.
\end{eqnarray}
This is sufficient for most of the higher partial waves. \item
One sign change and no bound state ($^1S_0^+,~^3P_0^+$ partial
states). In this case the energy-dependent expression for
$\lambda_l(s)$ is used (see \cite{Schwarz:1980bc} and references therein):
\begin{eqnarray}
\lambda_{ij}(s)=(s_0-s)\bar\lambda_{ij},
\end{eqnarray}
Here the parameter $s_0$ is introduced to reproduce the sign
change in the phase shifts at the position of the experimental
value for the kinetic energy \tlab\ where they are equal to zero.
It is added to the other parameters of the kernel.
\end{enumerate}

The calculation of the parameters is performed by using
Eqs.(\ref{T_norm_s}), (\ref{low}) and expressions given in two
previous sections to reproduce experimental values for all
available data from the SAID program (http://gwdac.phys.gwu.edu)
for the phase shifts. The the low-energy scattering parameters are
taken from~\cite{Dumbrajs:1983jd}.

\begin{figure}
\begin{center}
\includegraphics[width=95mm]{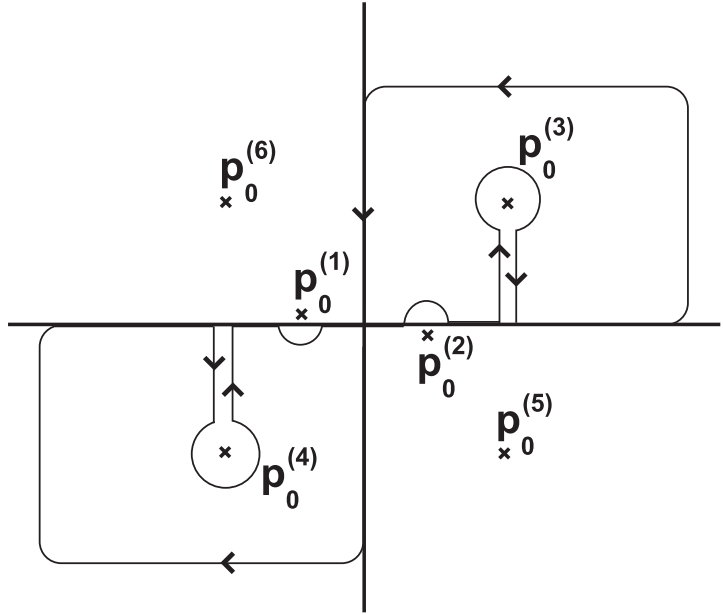}
\caption{{Contour for integration over $p_0$ after the Wick
rotation.}} \label{contour_w}
\end{center}
\end{figure}

The calculations are performed in two independent ways. The first
one is based on using the Cauchy theorem for the integration over
$p_0$ component of the nucleon four-momentum. The integration over
$|\p|$ is performed numerically. The second one is based on using
the Wick rotation \cite{Lee:1969fy}. In this case the contour of
integration Fig.\ref{contour} is deformed as depicted on
Fig.\ref{contour_w}. All integrals are calculated numerically with
the technique elaborated in the paper \cite{Fleischer:1975}.

Now we find the introduced parameters of the kernel:
\begin{enumerate}
\item
For $P$ waves the minimization procedure for the function
\begin{eqnarray}
\chi^2=\sum\limits_{i=1}^{n} (\delta^{\rm
exp}(s_i)-\delta(s_i))^2/(\Delta\delta^{\rm exp}(s_i))^2
\label{mini_p}
\end{eqnarray}
is used. Here $n$ is a number of available experimental points.
\item For the $^1S_0^+$ wave the values of the scattering length $a$
is also included into the minimization procedure
\begin{eqnarray}
\chi^2=\sum\limits_{i=1}^{n} (\delta^{\rm
exp}(s_i)-\delta(s_i))^2/(\Delta\delta^{\rm exp}(s_i))^2
+(a^{\rm exp}-a)^2/(\Delta a^{\rm exp})^2.
\label{mini_s}
\end{eqnarray}
\end{enumerate}
The effective range $r_0$
is calculated via the obtained
parameters and compared with the experimental value $r_0^{\rm exp}$.

The description of $P$ waves by the two-rank kernel is denoted by
MY2 for the first case of a relativization procedure; MYQ2, for
the second one. For the $^1S_0^+$ partial state the notation MY3
and MYQ3, respectively, are used.

The calculated parameters of the considered kernels are listed in
Tables~\ref{p_param} and \ref{1s0_param} (here the values of $s_0$
are presented, too). In Table \ref{lep}, the
calculated low-energy scattering parameters for the $^1S_0^+$ wave
are compared with their experimental values.

In Figs.\ref{1p1}-\ref{1s0}, the results of the phase
shift calculations are compared with experimental data
(the used notation is described in the following Section \ref{sect8}) and
two alternative descriptions by CD-Bonn \cite{Machleidt:2000ge} and SP07
\cite{Arndt:2007qn}. In the discussion of the $\Sp^+$ channel the nonrelativistic Graz II \cite{Mathelitsch:1981mr},
as an alternative model with a separable kernel, is also presented.
\begin{center}
\begin{table}
\caption{Parameters of the two-rank kernel with modified
Yamaguchi functions for $P$ waves.}
\centering
\begin{tabular}{l|ccc}
\hline\hline
           &          & MY2        &          \\
\hline
           &$^1P_1^+$ & $^3P_0^+$ & $^3P_1^+$ \\
\hline
$\bar\lambda_{11}^{{\phantom{1}}^{\phantom{1}}}$    (GeV$^4$) &  0.05412952  & -923.8881  &   0.06125619  \\
$\bar\lambda_{12}$    (GeV$^4$) &  1.925       & -102.0961  &   2.068215    \\
$\bar\lambda_{22}$    (GeV$^4$) &  7.975       &  4.346553  &   24.48148    \\
$\beta_{1}$           (GeV)     &  0.1244769   &  0.958602  &   0.1224502   \\
$\beta_{2}$           (GeV)     &  0.6228701   &  1.897255  &   0.5822389   \\
$\alpha_{1}$          (GeV)     &  0.2         &  0.759970  &   0.2107709   \\
$\alpha_{2}$          (GeV)     &  0.5984991   &  0.687087  &   0.5927882   \\
$p_{c2}$              (GeV$^2$) &  1.035       & -133.2385  &   0.9476951   \\
$s_0$                 (GeV$^2$) &              &  3.8682    &               \\
\hline\hline
           &         &  MYQ2     &          \\
\hline
           & $^1P_1^+$ & $^3P_0^+$ & $^3P_1^+$\\
\hline
$\bar\lambda_{11}^{{\phantom{1}}^{\phantom{1}}}$    (GeV$^4$)&  0.3086486 & -55.88270  &  0.07648584 \\
$\bar\lambda_{12}$    (GeV$^4$) &  1.606382  & -763.1649  &  1.567463   \\
$\bar\lambda_{22}$    (GeV$^4$) & -5.797411  & -5325.327  &  21.25497   \\
$\beta_{1}$           (GeV)     &  0.2150637 &  0.5008744 &  0.2109235  \\
$\beta_{2}$           (GeV)     &  0.9582849 &  2.4269161 &  0.4260685  \\
$\alpha_{1}$          (GeV)     &  0.2       &  0.6863803 &  0.2        \\
$\alpha_{2}$          (GeV)     &  0.2       &  0.2       &  0.2        \\
$p_{c2}$              (GeV$^2$) &  12.47736  &  5.866078  &  0.007380749\\
$s_0$                 (GeV$^2$) &            &  3.8682    &           \\
\hline\hline
\end{tabular}\label{p_param}
\end{table}
\end{center}
\begin{center}
\begin{table}
\caption{Parameters of the three-rank kernel with modified
Yamaguchi functions for $^1S_0^+$ state.}
\centering
\begin{tabular}{lcclc}
\hline\hline

                            &          & MY3  &                             &             \\
\hline $\bar\lambda_{11}^{{\phantom{1}}^{\phantom{1}}}$    (GeV$^2$) & -0.2922173 &      &
$\beta_{1}$       (GeV)      & 0.7018063 \\
$\bar\lambda_{12}$    (GeV$^2$) & -3.953624  &     &  $\beta_{2}$
(GeV) & 4.381178  \\
$\bar\lambda_{13}$    (GeV$^2$) &  1.035416  &     &  $\beta_{3}$
(GeV)      & 1.137604  \\
$\bar\lambda_{22}$    (GeV$^2$) & -11268.52  &     &  $\alpha_{1}$
(GeV)      & 1.297282  \\
$\bar\lambda_{23}$    (GeV$^2$) &  331.1130  &     &  $\alpha_{2}$
(GeV)      & 4.612956  \\
$\bar\lambda_{33}$    (GeV$^2$) &  70.37369  &     &  $\alpha_{3}$
(GeV)      & 0.6752485 \\
$s_0$             (GeV$^2$) &  4.0279    &     &  $p_{c1}$
(GeV$^2$) & 38.20462 \\
                            &            &     &  $p_{c2}$          (GeV$^2$)  & 34.53211  \\
\hline\hline
&          & MYQ3 &                             &             \\
\hline
$\bar\lambda_{11}^{{\phantom{1}}^{\phantom{1}}}$    (GeV$^2$) & 723.7399   &     &  $\beta_{1}$       (GeV)      & 2.463736  \\
$\bar\lambda_{12}$    (GeV$^2$) & 550.4827   &     &  $\beta_{2}$       (GeV)      & 1.266692 \\
$\bar\lambda_{13}$    (GeV$^2$) & -1583.031  &     &  $\beta_{3}$       (GeV)      & 7.714815  \\
$\bar\lambda_{22}$    (GeV$^2$) & 132.7836   &     &  $\alpha_{1}$      (GeV)      & 3.752405  \\
$\bar\lambda_{23}$    (GeV$^2$) & -14.26609  &     &  $\alpha_{2}$      (GeV)      & 2.117966  \\
$\bar\lambda_{33}$    (GeV$^2$) & -26005.23  &     &  $\alpha_{3}$      (GeV)      & 1.338336  \\
$s_0$             (GeV$^2$) &  4.0279    &     &  $p_{c1}$          (GeV$^2$)  & -44.84322 \\
                            &            &     &  $p_{c2}$          (GeV$^2$)  & 247.2476  \\

\hline\hline
\end{tabular}\label{1s0_param}
\end{table}
\end{center}
\begin{center}
\begin{table}
\caption{The low-energy scattering parameters and
for the singlet $(s)$ $^1S_0^+$ wave.} \centering
\begin{tabular}{lcc}
\hline\hline
           &$a_{s}$(fm) & $r_{0s}$(fm) \\
\hline
MY3        &-23.750      & 2.70         \\
MYQ3       &-23.754      & 2.78         \\
Experiment &-23.748(10)  & 2.75(5)      \\
\hline\hline
\end{tabular}\label{lep}
\end{table}
\end{center}
\section{Discussion}\label{sect8}
In this section, the review of the results of our calculations with
two methods of
relativization is performed.

\begin{figure}
\begin{center}
\includegraphics[width=135mm]{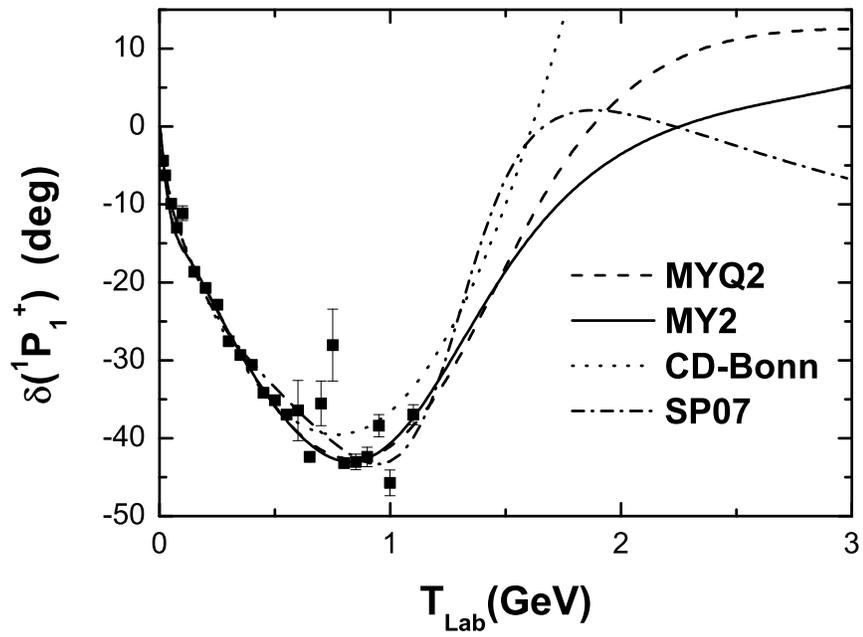}
\caption{{Phase shifts for the $^1P_1^+$ wave.}}
\label{1p1}
\end{center}
\end{figure}

In. Fig.\ref{1p1}, we can see that all of the calculations
including nonrelativistic CD-Bonn give a quite good description of the
$^1P_1^+$ state. In our previous works \cite{Bondarenko:2008fp,Bondarenko:2008ha}, it was shown that the one-rank kernel is already
sufficient to reproduce the phase shifts in this case. The reason is
the simplicity of their behavior and the narrowness of the energy
interval where they are known. So if the simplicity of the model
to perform calculations in the $^1P_1^+$ state is preferable, it is
sufficient to use the one-rank kernel. For the energies
$\tlab>1.1$\,GeV the resulting functions become very different. To
make choice in favour of one of the parametrizations, it is
necessary to have experimental data for larger energies.

\begin{figure}
\begin{center}
\includegraphics[width=135mm]{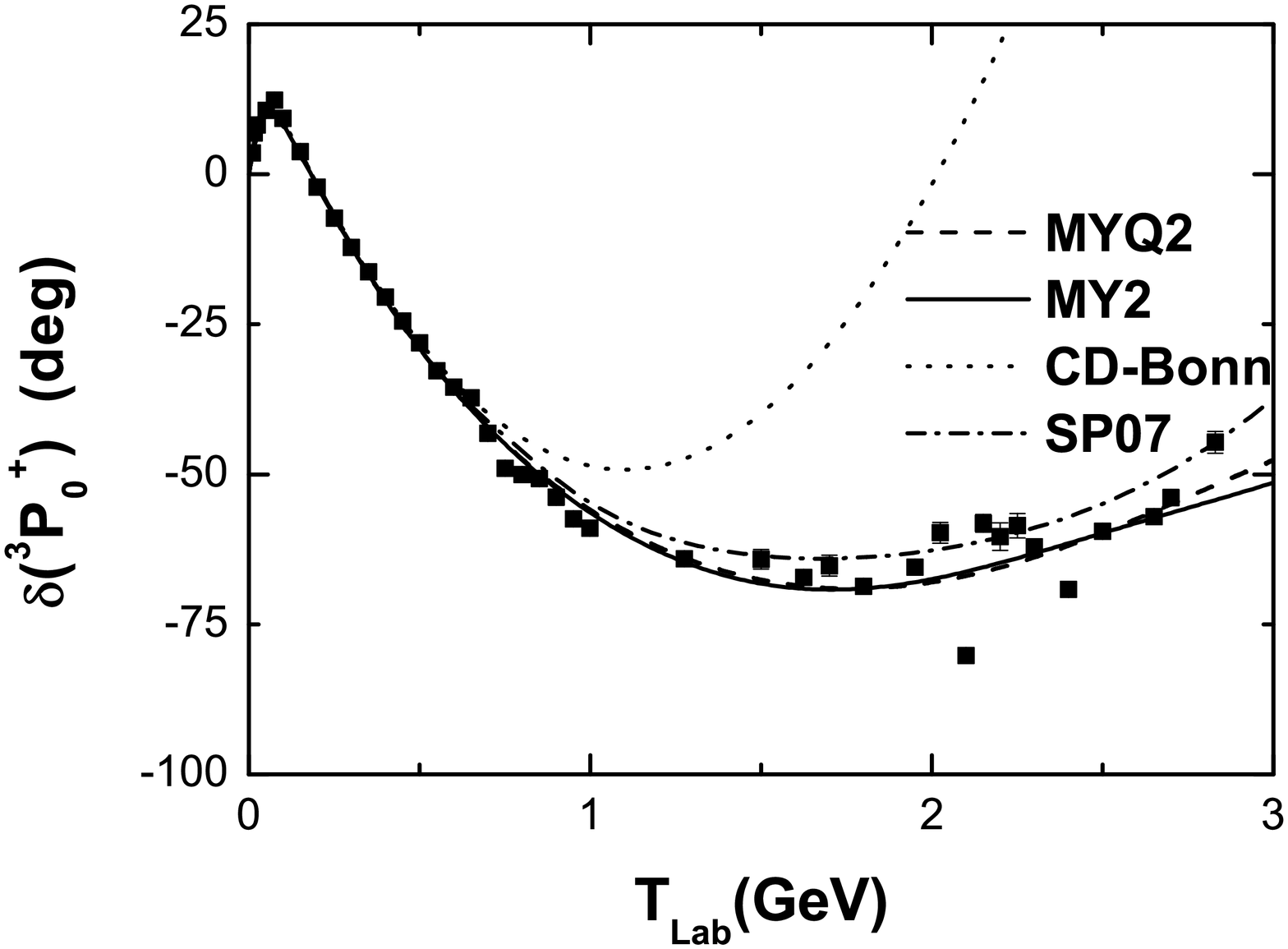}
\caption{{Phase shifts for the $^3P_0^+$ wave.}}
\label{3p0}
\end{center}
\end{figure}

In Fig.\ref{3p0}, the results of the calculations for the $^3P_0^+$
partial state are depicted. The comparison of our and other model
calculations demonstrates a reasonable agreement with the
experimental data in the whole range of energies except the
nonrelativistic CD-Bonn which works till $\tlab\sim 0.75$\,GeV.
The good description of the phase shifts in $^3P_0^+$ can also be
archived by using the one-rank interaction kernel with extended
Yamaguchi-type form factors (see \cite{Bondarenko:2008fp,Bondarenko:2008ha}). As in the previous case, the choice of the
kernel is defined, by reasons of convenience, as applied to a
specific problem.

\begin{figure}
\begin{center}
\includegraphics[width=135mm]{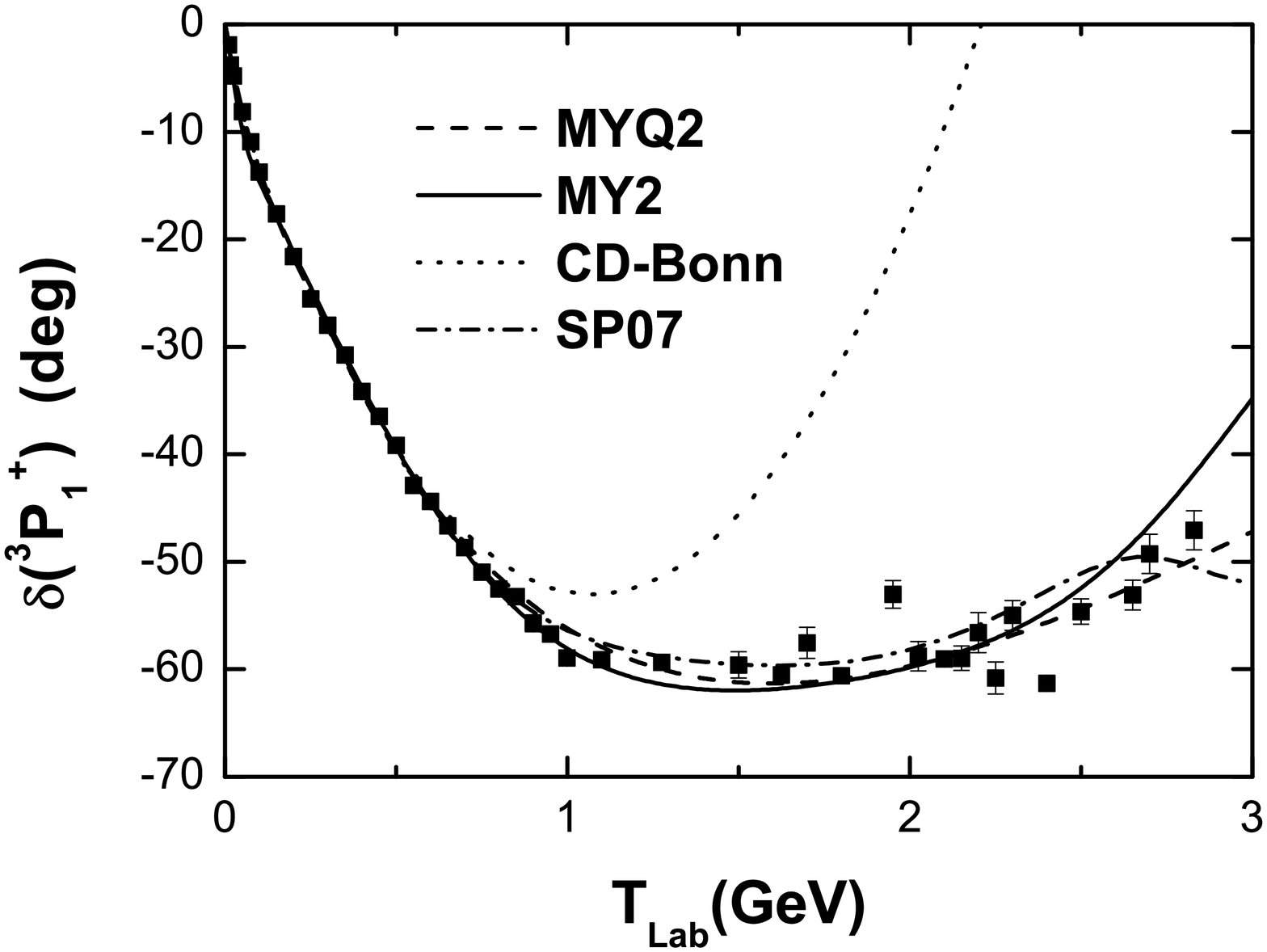}
\caption{{Phase shifts for the $^3P_1^+$ wave.}}
\label{3p1}
\end{center}
\end{figure}

The phase shift calculations for the $^3P_1^+$ partial state are
presented in Fig.\ref{3p1}. All models except CD-Bonn give various
but acceptable within the limits of error results. The CD-Bonn
works for $\tlab\leqslant0.66$\,GeV. As for our previous one-rank
kernel, the increase of the rank allows us to improve the
description and embrace all experimental data.

Our general conclusion about the description of $P$ states is that
using the two-rank kernel allows to reproduce the phase shifts in
good agreement with experimental data. If the simplicity is
the main requirement of performing calculations, the one-rank
kernel for $^1P_1^+$ and $^3P_0^+$ states and two-rank one for
$^3P_1^+$ can be used. Here we talk about phases in the whole energy
range for $\tlab$. However, if the use of a common interaction kernel
is preferable, then calculations with two-rank interaction
kernel should be done.

\begin{figure}
\begin{center}
\includegraphics[width=135mm]{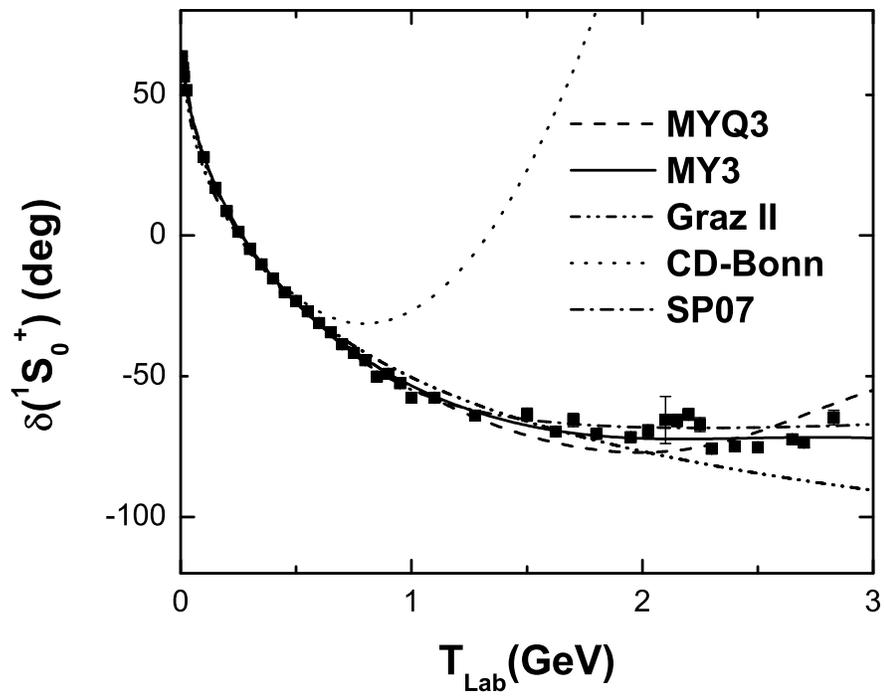}
\caption{{Phase shifts for the $^1S_0^+$ wave.}}
\label{1s0}
\end{center}
\end{figure}

From Fig.\ref{1s0}, where the results of calculations for $^1S_0^+$
are presented, it can be seen that CD-Bonn is quite good till
$\tlab\sim0.66$\,GeV. All the other results are in agreement with
measured phase shifts except Graz II which works for
$\tlab\leqslant2$\,GeV. We succeed in an acceptable description at the
cost of increasing the kernel's rank. Kernels of lower ranks
are proper only for energies $\tlab\leqslant$1\,GeV.

From the presented figures the similarity of the calculations with the
MY and MYQ form factors can be noted. Thus, as for the description
of phase shifts and low-energy parameters there is no difference
which variant of a relativistic generalization to choose. The
choice should be dictated by the convenience of performing
calculations within some
concrete problem.
\section{Conclusion}\label{sect9}
Using the multirank kernels (two-rank for $P$ waves, three-rank
for the $^1S_0^+$ partial state) we have constructed an adequate
description of all existent experimental data for phase shifts
taken from SAID and low-energy parameters with capable accuracy.

The results for two different methods of a relativistic
generalization of initially nonrelativistic Yamaguchi-type
form factors were considered. As it was shown they
lead to slightly different descriptions of phases and low-energy
parameters. Hence, the choice of the concrete form of functions for
performing calculations of any process is governed only by convenience.

In spite of the fact that the model functions have a simple form
there are quite a few parameters in the description of the data. This is necessitated by introduction of an additional
parameter $\alpha$ so that integrands
containing form factors of the separable kernel could not have poles. In
particular, using this type of kernels will make the numerical
calculations of the electrodisintegration far from the threshold
possible without resorting quasipotential or nonrelativistic
models.
\section{Acknowledgements}\label{sect10}
We are grateful to Drs. A.A. Goy and D.V. Shulga for their interest in
this work. We would also like to thank Dr. Y. Yanev for computational
advice and Professor G. Rupp for his leading questions and
recommendations for a more precise formulation of the basis of our
model.

\end{document}